\magnification=\magstep1
\hsize 6.0 true in
\vsize 9.0 true in
\voffset=-.5truein
\pretolerance=10000
\baselineskip=22truept

\font\tentworm=cmr10 scaled \magstep2
\font\tentwobf=cmbx10 scaled \magstep2

\font\tenonerm=cmr10 scaled \magstep1 
\font\tenonebf=cmbx10 scaled \magstep1

\font\eightrm=cmr8
\font\eightit=cmti8
\font\eightbf=cmbx8
\font\eightsl=cmsl8
\font\sevensy=cmsy7
\font\sevenm=cmmi7

\font\twelverm=cmr12  
\font\twelvebf=cmbx12
\def\subsection #1\par{\noindent {\bf #1} \noindent \rm}

\def\mid {\let\rm=\tenonerm \let\bf=\tenonebf \rm \bf}

\def\para{\par \vskip 12 pt}

\def\head{\let\rm=\tentworm \let\bf=\tentwobf \rm \bf}

\def\heading #1 #2\par{\centerline {\head #1} \smallskip
 \centerline {\head #2} \vskip .15 pt \rm}

\def\eight{\let\rm=\eightrm \let\it=\eightit \let\bf=\eightbf 
\let\sl=\eightsl \let\sy=\sevensy \let\m=\sevenm \rm}

\def\foots{\noindent \eight \baselineskip=10 true pt \noindent \rm}
\def\sexion{\let\rm=\twelverm \let\bf=\twelvebf \rm \bf}

\def\section #1 #2\par{\vskip 20 pt \noindent {\mid #1} \enspace {\mid #2} 
  \para \noindent \rm}

\def\abstract#1\par{\para \foots {\bf Abstract: \enspace}#1 \para}

\def\author#1\par{\centerline {#1} \vskip 0.1 true in \rm}

\def\abstract#1\par{\noindent {\bf Abstract: }#1 \vskip 0.5 true in \rm}

\def\sqr#1#2{{\vcenter{\vbox{\hrule height.#2pt
  \hbox {\vrule width.#2pt height#1pt \kern#1pt
  \vrule width.#2pt}
  \hrule height.#2pt}}}}

\def\n{\noindent}
\def\s{\smallskip}
\def\m{\medskip}
\def\b{\bigskip}
\def\c{\centerline}

\def\gne #1 #2{\ \vphantom{S}^{\raise-0.5pt\hbox{$\scriptstyle #1$}}_
{\raise0.5pt \hbox{$\scriptstyle #2$}}}

\def\ooo #1 #2{\vphantom{S}^{\raise-0.5pt\hbox{$\scriptstyle #1$}}_
{\raise0.5pt \hbox{$\scriptstyle #2$}}}


\n~~~~~~~~~~~~~~~~~~~~~~~~~~~~~~~~~~~~~~~~~~~~~~~~~~IUCAA-33/97
\c{\bf\mid On the Schwarzschild field}
\b
\b
\b
\b
\b 
\b
\b
\c{\bf Naresh Dadhich\footnote{$^* $}{E-mail : nkd@iucaa.ernet.in}}
\c{\bf Inter University  Centre for Astronomy \& Astrophysics}
\c{\bf P.O. Box 4, Pune-411007, India}
\b
\b
\b
\b
\b
\baselineskip=16truept
\c{\bf Abstract}
\s 

\n General   relativity   is  a  non-linear   theory   with   the 
distinguishing  feature  that   gravitational field  energy  also 
acts  as  gravitational  charge    density.  In  the   well-known 
Schwarzschild  solution describing field of an  isolated  massive 
body  at rest, the scalar function $\phi$ characterising the field  acts 
as  a gravitational potential as well as it curves space part  of 
spacetime.  We  demonstrate  explicitly that  it  is  the  latter 
property  that accounts for the non-linear ( gravity as  its  own 
source ) aspect which is not explicit in usual derivations. It is 
worth noting that the Einstein vacuum equations ultimately reduce 
to  the Laplace equation and its first integral which fixes  zero 
of $\phi$ at infinity. Thus the Schwarzschild field 
alongwith  its asymptotic flat character is completely determined 
without  application  of 
any   boundary  condition by  the  field 
equations  themselves. That means non-zero constant value of $\phi$ 
will have non-vacuous effect. It in fact produces 
stresses  exactly  of  the form required to  represent  a  global 
monopole.   By  retaining  freedom  of  choosing  zero   of  $\phi$,
which will break asymptotic flatness, we can obtain the 
Schwarzschild  black hole with global monopole charge. It is  the 
non-linear  aspect responsible for ``curving'' space, which has  no 
Newtonian analogue, survives even when $\phi$ is constant  but 
not zero.

\b
\b
\b
\b
\n PACS numbers : 0420, 9880.

\vfill\eject

\baselineskip=22truept
\item{\bf 1.} {\bf Introduction}
\s
\n In  the  Newtonian theory (NT), gravitational field  is  entirely 
determined  by the equation $\bigtriangledown^2 \phi = 4 \pi \rho$
and motion of particle in  the 
field  is given by $\ddot r = - \bigtriangledown \phi$.
Here we have set $G  =  1$, 
and $\rho$ is the matter density. In absence of matter $\rho = 0$,
we  have 
the  Laplace equation, which for a radially symmetric  field  has 
the  well-known  solution $\phi = k - {M/r}$. It  describes  gravitational 
field of a body of mass $M$ at rest at the origin of the coordinate 
system. The  arbitrary constant $k$ can be chosen freely to set  the 
zero  of the potential $\phi$. Normally one sets it to zero  to  have 
it  vanish  at infinity. The main criticism  of  the  above 
equation  is, one it does not prescribe propagation of  field  in 
space  and  two it cannot take into account  the  distinguishing 
feature  of gravity, gravitational field being its own  source. To 
incorporate  these  aspects  we  will  have  to  go  to   general 
relativity (GR).
\s
\s
\n The  latter property would demand that in matter free region  the 
equation should be modified to

$$ \bigtriangledown^2 \phi = {1 \over 2} (\bigtriangledown \phi)^2 \eqno (1.1)$$
 
\n to bring  in  contribution  of  gravitational  field  energy  to 
gravitational charge density. Let us consider the field of a  mass 
point  in  GR  to  see  whether do  we  really  solve  the  above 
equation? It  turns out that we still solve the  Laplace  equation 
rather  than  eqn.(1.1). Then  how  does  GR  take  care  of  this 
aspect? Eqn.(1.1) would be referred to flat spacetime while in  GR 
spacetime  is  curved and it is the curvature of  spacetime  that 
describes  gravitational field. That opens up an interesting
possibility  that 
one can retain the Laplace equation and non-linear aspect of  the 
field (being its own source) is taken care by  curvature  of 
space. It is remarkable that this is what exactly happens.
\s
\s
\n One of the main aims of this paper is to demonstrate this  aspect 
explicitly   and  clearly,  which  is  generally   neither   duly 
emphasized  nor  appreciated  in  the  usual  derivation  of  the 
Schwarzschild  solution. Secondly  GR field equations  for  vacuum 
ultimately  reduce to two equations of which one is the  good 
old  Laplace equation and the other is its first integral,  which 
determines the free parameter  $k = 0$ [1]. Thus asymptotic flatness  of 
the Schwarzschild field is not due to a proper choice of boundary 
condition  but  instead  is  entirely  determined  by  the  field 
equations  themselves. We have no choice to make $\phi$ zero  anywhere 
else than infinity. This is what is reflected in the fact that the 
Schwarzschild solution is the unique spherically symmetric vacuum 
solution. This aspect though known [2,3] but not well-known enough 
and certainly not without surprise for many. It is well-known that 
the  Schwarzschild  spacetime is fully written in  terms  of  the 
Newtonian potential. In both the theories there is only one scalar 
function that describes the field and hence it should arouse some 
surprise  that  GR  equations  determine  this  scalar   function 
absolutely  while NT offers  freedom of additive constant to  fix 
its  value  at the boundary. It is to be noted that  GR  does  not 
offer  this  freedom. 
\s 
\s  
\n In  the  next  Section we shall first  derive  the  Schwarzschild 
solution explicitly demonstrating, how field equations  determine 
the  field entirely without reference to any  boundary  condition  
and how non-linearity is taken care  by ``curving'' space  part 
of  the  metric. It  would be shown that this  part  of  curvature 
survives  even when the scalar function determining the field  is 
constant but not zero. In Section 3 we shall argue that the  basic 
character  of  the  Schwarzschild field  will  not  be  disturbed 
significantly even if we retain the free parameter $k$. As is  clear 
that  vacuum equations will then no longer remain   satisfied. The 
metric  will neither be asymptotically flat. In principle the 
Schwarzschild solution should not be asymptotically flat to permit
presence of other matter-energy in the Universe. Realistically
we should break asymptotic flatness.  This is the only and perhaps  the 
most  harmless way of rendering asymptotic non-flat character  to 
the   Schwarzschild  field. By breaking asymptotic  flatness
we  give  a  generalization   of   the 
Schwarzschild  solution  which is equivalent to adding  a  global 
monopole  to  it  [1,4]. Its effect on  particle  orbits  and  the 
Hawking  radiation  has  been  studied  recently  [5]. Finally  we 
conclude  with a discussion raising some points of principle  and 
concept.
\vfill\eject
\item{\bf 2.}{\bf Schwarzschild's solution }
\s
\n First we shall derive the Schwarzschild spacetime explicitly
demonstrating how non-linear aspect of gravity goes into curving
space and there is no scope for application of any boundary
condition. Asymptotic flatness of the solution
is implied by the equations themselves. We shall carry out
the analysis in curvature, Kerr-Schild and isotropic coordinates
so as to indicate that it is not a coordinate dependent result.
\s
\s
\item{\it 2.1} {\it The curvature coordinates :}
\s
\n Let us begin with 

$$  ds^2  = B dt^2 - A dr^2 - r^2 ( d \theta^2 + sin^2  \theta  d 
\varphi^2) \eqno (2.1) $$

\n where $ A $ and $B $ are functions of $r $ and $t $. Note that
the radial coordinate $r $ defines the area of 2-sphere and hence
it has proper physical meaning.
The Ricci tensor for it reads as follows:

$$  R^0_0  =  -  {1 \over 2 AB}  \bigg[  \bigtriangledown^2  B  - 
{B^{\prime}  \over 2} ({A^{\prime} \over A} +  {B^{\prime}  \over 
B})  - \ddot A + {\dot A^2 \over 2A} + {\dot A \dot B  \over  2B} 
\bigg]  \eqno (2.2) $$

$$  R^1_1  =  R^0_0  +  {1  \over  Ar}  ({A^{\prime}  \over  A}  + 
{B^{\prime} \over B}) \eqno (2.3) $$

$$ R^2_2 = R^3_3 = {1 \over r^2} \bigg[1 - {1 \over A} + {r \over  2A} ({A^{\prime} \over A} - {B^{\prime} \over B}) \bigg] \eqno (2.4) $$

$$ R_{01} = -{\dot A \over Ar} \eqno (2.5) $$

\n where $\bigtriangledown^2 $ is the Euclidian Laplacian, and 
an overhead prime and dot denote derivative w.r.t.~$r $ and $t $.
\s
\s
\n From physical point of view it is obvious that field of a  static 
body cannot depend upon $t $ which is ensured by $R_{01} = 0 $ in (2.5)
which implies $A = A(r)$.
It is well-known that geodesic equation with $B = 1 + 2 \phi $ and 
$A = 1 $(space part of the metric being flat) incorporates the  Newtonian 
equation  of  motion $\ddot r = - \bigtriangledown \phi $ for slow motion
and weak field. Then $R^0_0 = 0$ will take the form (1.1), taking into
account the red-shift factor for the field energy density. 
This indicates clearly the appearance
of field energy density as gravitational charge density.
If we  now  ``curve''  the space part by  
introducing  $A $, then  demanding $R_{01} = 0, R^0_0 = R^1_1 $ 
from (2.2) and (2.3) that give

$$ \dot A = 0, {A^{\prime} \over A} + {B^{\prime} \over B}  = 
0 \Rightarrow  AB = f(t) = 1. \eqno (2.6) $$

\n Note that this reduces $R^0_0 = 0$ to the Laplace equation indicating
non-linear aspect is taken care of by curvature of space. Introduction
of $A$ means curving space part  of the metric, which cancels out contribution
of field energy density from gravitational charge density leading
to the Laplace equation again. It is this point
which we wish to emphasise. Else one does not see how non-linear aspect
of gravity is accounted for in the Schwarzschild solution.
\s
\s
\n Here $f(t) $ is absorbed by redefining $t $ which neither amounts 
to any  loss of generality  nor of invoking any boundary  condition [2,3].
In the usual derivation it is here the asymptotic flatness is invoked
to set $f(t) = 1 $. Even if $f(t) $ is retained the field will be
asymptotically flat and the solution with $f \not= 1 $ is physically
indistinguishable from the one with $f = 1 $. So we apply no
boundary condition to set $f = 1 $. It only identifies the coordinate
$t $ with that of the asymptotic observer.
Hence $B = A^{-1} = 1 + 2 \phi (r) $, say (Birkhoff's theorem).
Using  this  in (2.2)  and (2.4) we are finally lead to the 
following two linear equations [1], 

$$  R^0_0  =  -\bigtriangledown^2  \phi  =  -  {1  \over  r}   (r 
\phi)^{\prime \prime} = 0 \eqno (2.7) $$

$$ R^2_2 = -{2 \over r^2} (r \phi)^{\prime} = 0. \eqno (2.8) $$
\s

\n 
Thus we have come back to the good old Laplace equation. 
Note that (2.8) is the first integral of (2.7) and hence we just need
to integrate (2.8) to get to the Schwarzschild solution.As 
envisioned earlier the curvature of space (i.e. $A \not= 1 $ and
using (2.6) in (2.2)) just exactly cancels out
the field energy density term in (2.2) and synthesises its effect
in the geometry of space. The former equation admits the well-known
general solution

$$ \phi = k - {M \over r} \eqno (2.9) $$

\n while  the latter determines $k = 0 $. Thus we obtain the 
Schwarzschild solution.
\s
\s
\n The important point  to 
note  is that we had no freedom to use any boundary  condition, in 
particular  asymptotic flat behaviour of the solution is implied by the  field 
equations themselves. That is the solution is fully determined by the theory 
leaving  no scope for boundary conditions. Secondly it is very 
insightful to see how non-linear aspect is incorporated by ``curving''
space. These points are not brought forth emphatically in the usual
derivations.
\s
\s
\n Note  that  eqns. (2.7) and (2.8) are the exact  Einstein's 
 equations  and $\phi $ satisfies  the  Laplace  equation (which
is $R^0_0 = 0 $) without any approximation. Since it is
$R^0_0 = 0 $ that is supposed to be the analogue of the Newtonian
Laplace equation which defines gravitational potential, hence $\phi $
in (2.7) would define relativistic gravitational potential (an analogue
of the Newtonian potential taken over to GR) exactly at least in the 
coordinates used without any approximation 
of weak  field and  slow motion. The first point we wish 
to make is  that  unlike 
the   Newtonian  theory, GR  determines  scalar function $\phi$ describing
the field 
absolutely. That  means $\phi = k \not= 0 $ is not a  solution  of  
Einstein's equation $R_{ik} = 0 $ as is clear from (2.8) and 
it produces  non-zero   curvature. Thus  constant  $\phi$  
attains   non-trivial physical meaning. This is a very curious 
and unique feature of  GR. 
\s
\s
\n At this stage it may be instructive to have a look at the
Riemann curvatures of the metric (2.1) with $B = A^{-1} = 1 + 2 \phi (r),$

$$R^{01}_{~~01} = \phi^{\prime \prime}, ~R^{02}_{~~02} = R^{12}_{~~12}
= {\phi^{\prime} \over r}, ~ R^{23}_{~~23} = {2 \phi \over r^2}. \eqno (2.10) $$

\n It is clear that $\phi = const.$ does not make $R_{2323} = 0.$ This
is because $\phi$ appears as it is in curvature imparting physical
meaning to itself. This is in contrast to usual concept of potential
in classical physics. It is purely a relativistic feature arising
from non-linearity of the theory.
\s
\s
\n Recall  that we have argued above that $B $ was responsible 
for  the Newtonian  acceleration in the geodesic equation 
while $A $  brought in the non-linear effect, gravity as its 
own source. When $\phi $  is constant, both   $A $  and $ B $ 
 are  constants. Clearly   gravitational 
acceleration  vanishes and $B $ can be absorbed by redefining  $t 
$. But $A $ 
which  represents non-linear aspect cannot be got rid of unless
$\phi = 0$, and  it 
persists  even when $\phi $ is constant. Since the Newtonian theory  was 
free of the non-linear aspect and space was flat, 
that is why constant potential  was 
physically  inert. But it is not so in GR for relativistic potential 
$\phi$, an analogue
of the Newtonian potential, produces curvature
in space. This is the crucial point that in GR, $\phi $ not only
determines field and motion of test particles but also curves space.
The latter
property is sensitive to the absolute value of $\phi $. This
is how $\phi $ attains physical meaning. In GR the field equations 
not only  regulate  the behaviour  of the field but they also determine  
the spacetime itself. Hence they provide a more constrained system
than that in NT.
\s
\s
\n It is well-known that metric potentials $A$ and $B$ contribute
equally in the Schwarzschild solution for bending of light ray,
that is half the contribution comes from gravitational potential,
$B$ and the half from curvature of space through $A$, caused
by non-linear aspect of the field. To gain some more insight
into their working let us consider their effects separately.
\s
\s
\item{(i)} Let $B = 1 - 2M/r$ and $A = 1$, which will generate
the stresses,

$$ T^0_0 = 0, 8 \pi T^1_1 = - {2 M/r^3 \over 1 - 2M/r}, 8 \pi T^2_2 =
{(M/r^3) (1 - M/r) \over (1 - 2M/r)^2}. $$

\n Note that energy density $T^0_0 = 0$ but gravitational
charge density, $4 \pi \rho_c = (M^2/r^4)(1 - 2M/r)^{-2} \not=0$.
The proper red shifted acceleration of a free particle relative
to infinity is given by $\alpha \bigtriangledown (ln \alpha)$ where
$\alpha = B^{1/2}$, the lapse function, which will give
the Newtonian acceleration $M/r^2$. It is $\rho_c$ that is
responsible for this acceleration. Note that vanishing of
proper acceleration implies vanishing of $\rho_c$. The metric
potential therefore produces accleration as well as tidal acceleration
for radial and non-radial motion.
\s
\s
\item{(ii)} When $B = 1$ and $A = (1 - 2M/r)^{-1}$, we have

$$T^0_0 = 0 = T^0_0 - T^{\alpha}_{\alpha} = \rho_c, T^1_1 = M/4 \pi r^3,
T = 0. $$

\n Here not only energy density $T^0_0$ but $\rho_c$ and $T$
also vanish. Since $B=1$, which means vanishing of acceleration
for free particles as well as gravitational charge density
$\rho_c$. Radially falling particles will experience neither
acceleration nor tidal acceleration. The curvature of space
produced by $A$, will menifest only in tidal acceleration for
non-radial motion.
\s
\s
\n Roughly speaking $B$ accounts for the usual Newtonian gravity while
$A$ brings in the Einsteinian aspect of non-linearity of gravitational
field. When the two are synthesized together, then the Schwarzschild
solution follows.
\s
\s
\item{\it 2.2} {\it The Kerr-Schild coordinates :}
\s
\n This is the another coordinate system in which like the 
curvature coordinates 
the gravitational potential can be defined exactly. Here we  write 
the metric in the form,

$$ ds^2 = dt^2 - dr^2 - r^2 (d \theta^2 + sin^2 \theta d \varphi^2) +
2 \phi (dt + dr)^2 \eqno (2.11) $$

\n with

$$ \bigtriangledown^2 \phi = 0 \eqno (2.12) $$

\n which admits the general solution as given in (2.9). However  the 
vacuum  equation $R_{ik} = 0 $  will again demand $k =  0 $,  
which  implies asymptotic flatness (the Schwarzschild solution).
Note that  here 
again we have the Laplace equation for $R^0_0 = 0 $. 
It is legitimate to ask  what 
does  the classically trivial  solution  of  this  equation, $\phi = const. $
correspond to?
\s
\vfill\eject
\item{\it 2.3} {\it The isotropic coordinates :}
\s
\n At first sight it may appear that we can just transform from 
curvature to isotropic coordinates. The important point to note is that 
$k $ plays  a  non-trivial role in  the  transformation;  i.e.  the 
character of transformed metric becomes radically different  from 
the one when  $k $ is zero. In contrast to the above two cases,  there  does 
not  exist  a natural choice for potential in this case.  In  the 
above  cases, it was $R^0_0 = 0 $ that could be written as the  Laplace 
equation  while here it happens for $G^0_0 \equiv R^0_0 - {1 \over 2} 
~ R = 0 $. But $R^0_0 = 0 $ cannot be cast as the Laplace equation
without approximation. In the Newtonian limits of weak field and slow
motion, $R^0_0 \simeq G^0_0 $. However if we take the special
relativistic limit for which field is weak but motion is
relativistic, then $R^0_0 \not\simeq G^0_0 $ and it is $R^0_0 = 0 $
that should approximate to the Laplace equation and define potential. 
Here this does not happen and hence we cannot define the analogue
of $\phi $ except writing it in the isotropic coordinates. This is
because the isotropic $r $ is not the physical radial coordinate as
it does not define correctly the area of a sphere of radius $r $.
\s
\s
\n By  writing  the Ricci tensor in the  isotropic  coordinates  and 
solving the vacuum equations we shall once again demonstrate  that 
the solution fixes one of the free parameters which is equivalent 
to $k = 0 $ and that implies asymptotic flatness. We write the 
metric in the isotropic form as

$$ ds^2 = c^2 dt^2 - a^2 (dr^2 + r^2 d \theta^2 + r^2 sin^2 \theta
d \varphi^2). \eqno (2.13) $$

\n We  shall now consider the metric functions to be functions
of  $r $ alone, because the validity of the Birkhoff's
theorem is not in question. The non-zero Ricci components
 read as follows: 

$$ R^0_0 = -{1 \over a^2} \bigg[{c^{\prime \prime} \over c} +
{c^{\prime} \over c} \bigg( {a^{\prime} \over a} + {2 \over r}
\bigg) \bigg] \eqno (2.14) $$

$$ R^1_1 = -{1 \over a^2} \bigg[ {2a^{\prime \prime} \over a}
- {2a^{\prime^2} \over a^2} + { 2a^{\prime} \over ar} +
{c^{\prime \prime} \over c} - {a^{\prime} 
c^{\prime} \over ac} \bigg] \eqno (2.15) $$

$$ R^2_2 = -{1 \over a^2} \bigg[{a^{\prime \prime} \over a} 
+ {c^{\prime} \over c} \bigg({a^{\prime} \over a} + {1 \over r} 
\bigg) + {3a^{\prime} \over ar} \bigg]. \eqno (2.16) $$

\n The  first two equations  for $R^0_0 = R^1_1 = 0 $ 
yield  the  following 
first  integral  and  a  second  order  non-linear   differential 
equation, 

$$ 1 + {a^{\prime} \over a} r = k_1 c \eqno (2.17) $$

$$ \bigg({a^{\prime} \over a} r \bigg)^{\prime} ar^2 = k_1 k_2.
\eqno (2.18) $$   

\n This equation admits the general solution as given by

$$ a = r^n \bigg(1 + {M \over 2r^{n+1}} \bigg)^2 \eqno (2.19) $$

\n which in view of (2.17) fully determines the metric by giving 

$$ c = {1 - M/2 r^{n+1} \over 1 + M/2 r^{n+1}} \eqno (2.20) $$
  
\n where $M = k_2/k_1 $ and $n = k_1 - 1 $.
\s
\s
\n So far we have not used $ R^2_2 = 0 $, which had previously 
fixed $k = 0 $, it 
will now determine the arbitrary constant, $n = 0,-2, $ from 
 
$$R^2_2 = -~ {n(n+2) \over a^2 r^2}. \eqno (2.21) $$
 
\n Both  these values however yield the same spacetime. Hence  
fixing $n $ here is equivalent to fixing $k $  
in the curvature coordinates 
and the two are related as $2k = n(n+2) $. As  a 
matter  of  fact it can be verified that if we transform  from  the 
curvature  coordinates  to the isotropic  coordinates  with $k \not= 0, $ 
 we shall end up with (2.19) and (2.20). The constant $\phi $ in 
the curvature coordinates reflects in  a 
very different way in the  isotropic coordinates. This is because,
here the true radial coordinate (the one that defines the area
of a sphere) is $a r $ and not $r $ itself. Clearly 
in  the isotropic coordinates,  it  is  not 
obvious  to see the association of the parameter $n $ with  constant 
potential. It is however true that its being different from  zero 
and $-2$ only makes $R^2_2 $ alone non-zero. 
\s
\s
\n Let us see how $\phi $ is expressed in terms of the isotropic $r $,
$$ 2 \phi = n(n + 2) - {2M/r^{n + 1} \over (1 + M/2 r^{n+1})^2} 
\eqno (2.22) $$

\n and then we can also write

$$ a = {2M/r \over n(n+2) - 2 \phi}, ~c = [1 - n(n+2) + 2 \phi]^{1/2}. 
\eqno (2.23) $$

\n Note that $\phi = const. $ does not make $a = const. $, and
hence spacetime does not become flat. Obviously $\phi $ does not
satisfy the Laplace equation in these coordinates.
\s
\s
\n In   the   curvature  and  the   Kerr-Schild   coordinates,   the 
Schwarzschild  solution can be written in terms of the   
potential which arises as the solution of $\bigtriangledown^2 \phi = 0 $
corresponding to $R^0_0 = 0 $,  and  putting  that equal to 
constant  will  give  the 
spacetime   corresponding   to  $\phi = k $ in  (2.9). 
It is however not so transparent 
in the isotropic coordinates because the isotropic $r $ is not
the proper radial coordinate. This is also reflected in the fact
that in  the  former  cases  (2.9)  was  the  solution   of                   $R^0_0 = 0 $,  which  could be written as the Laplace equation 
by using $R^0_0 = 0 $ and $R^1_1 = 0 $, while  in  the 
latter  case   this doesn't happen (what does take  the  required 
form  is $G^0_0 = 0 $). This is why in the isotropic  coordinates  the 
point  we wish to raise does not become visible directly but  one 
has  to go through the derivation  of the Schwarzschild  solution 
in  these  coordinates  to  see  how  the  free  parameter  
$n  $ is determined. 
\s
\s
\n The point that emerges without any ambiguity from the 
above  discussion  is  that  asymptotic  flat  character  of  the 
Schwarzschild  solution  is  dictated  by  the  field   equations 
themselves  leaving  no  scope  for  any  boundary  condition  to 
operate. The choice different from $k = 0 $ or $n = 0, -2 $ implies
$R^2_2 \not= 0 $ and asymptotic non-flatness.  This  means  the  
field  equations  cannot 
admit an asymptotically  non-flat spherically symmetric 
vacuum  solution. This  is exactly what  the  uniqueness  of  the  
Schwarzschild  solution 
signifies.  The  amazing  thing  is that we  have  no  choice  to 
impliment  any  boundary  condition. 
\s
\s
\n In NT  we  had  the  general 
solution  given by (2.9), where we were free to choose $k $ to  fix 
zero  of  the  solution. This is a different  question  that  the 
canonical choice is $\phi = 0 $ at infinity, but we had the freedom  to 
make  another choice, which is not available in GR. Now the question
arises what happens if we have the true analogue of the Newtonian situation
with freedom to choose the constant $k$ ? This is what we take up
in the next Section. 
\b          
\item{\bf 3.} {\bf The generalized Schwarzschild field and global monopole }
\s
\s
\n The main question we wish to address is what happens if we do not 
let  $k = 0 $ in (2.9) (which has no physical effect in NT). In  that 
case $R^2_2 = -2k/r^2 \not= 0 $, however all other $R_{ik} $ 
will be  zero, 
and the  spacetime will not be empty. Let us ask  how  much  the 
empty space character of spacetime will be disturbed by this?  In 
NT vanishing of matter density indicates emptiness. Its  analogue 
in GR is the  gravitational  charge density defined by               

$$ - 4 \pi \rho_c = R_{ik} u^i u^k \eqno (3.1) $$

\n where $u^i $ is the timelike unit 4-velocity. That does however vanish 
even when $k \not= 0$. Clearly the parameter $k $ will 
not  have any physical effect at the Newtonian level. Though  the 
spacetime  is not strictly empty in the GR sense but it is  empty 
enough  in  the Newtonian sense because $\bigtriangledown^2 \phi = 0$. 
The other way of looking  at  it 
would  be  to  see  what happens  when  stresses  producing  zero       
$\rho_c $ are added to the Schwarzschild spacetime. That is all but
$R^2_2 = - 2k/r^2 \not= 0 $ vanish. That is a global monopole is added
to a Schwarzschild field [4]. It may be noted that $\rho_c = 0$ indicates
zero gravitational charge (mass).
\s
\s
\n The generalized Schwarzschild solution will read as

$$ds^2 = (1 + 2k - {2M \over r}) dt^2 (1 + 2k - {2M \over r})^{-1} dr^2
- r^2 (d \theta^2 + sin^2 \theta d \varphi^2) \eqno (3.2) $$                                                              
\n because $B = A^{-1} = 1 + 2 \phi, ~ \phi = k - {M / r}$. Similarly
the Reissner-N$\ddot o$rdstrom and the de Sitter spacetime can be generalized
by writing $\phi $ in (2.9) as $\phi = k - M/r + Q^2/2r^2 + \wedge r^2/6$.
\s
\s
\n It is clear that parameter $k$ has no effect on radial acceleration
for free particles. Hence the metric (3.2)
is very nearly equivalent to the Schwarzschild field. Particle orbits
 and the Hawking radiation have been examined to study the 
effect of $k$ [5]. It turns out that existence, boundedness and stability
of circular orbits scale up by the factor $(1+2k)^{-1}$, and the
perihelion and the light bending by $(1 + 2k)^{-3/2}$ while
the Hawking temperature scales down by $(1+2k)^2$ for a negative
$k$.
The  presence  of $ k $ will only be felt  when  we  consider 
geodesic deviation for transverse motion along $\theta -$ or $\varphi -$ direction. 
\s
\s
\n The metric gives rise to the stress system,
$$ 4 \pi T^0_0 = - {k \over r^2} = 4 \pi T^1_1 \eqno (3.3) $$

\n which  exhibits tension in the radial direction is  equal  to 
energy density and the transverse stresses being zero. This 
is   precisely  the  prescription  for a global monopole with
$k = - 4 \pi \eta^2$, where $\eta$ is the monopole charge [4]. This
exotic  object  is 
supposed  to  occur when global $0(3)$ symmetry is spontaneously
broken into $U(1)$ in phase transition in the  very  early 
Universe.  It  is  interesting  that  the  mandane  situation  of 
``constant $\phi$'' shares the spacetime description with such
an exotic object. 
\s
\s
\n Further the spacetime (3.2) is not asymptotically flat instead
it goes over to

$$ ds^2 = dt^2 - (1 + 2k)^{-1} dr^2 - r^2 (d \theta^2 + sin^2 \theta
d \varphi^2). \eqno (3.4) $$    

\n where $t$ has been redefined to absorb the factor $(1 + 2k).$ 
This  is indistinguishable from flat spacetime in all respects 
except for tidal acceleration for transverse motion. It produces
the monopole stresses (3.3) which have vanishing gravitational
charge density $\rho_c$.
For the metric (3.4), only space part is  curved 
and  its  curvature  at  a given $r (R^{23}_{~~23} = 2k/r^2 ) $
is  in  fact 
propotional  to  that  of a sphere of radius $r $. This is the only one
non-zero curvature component which  is
an invariant for spherical symmetry preserving 
coordinate transformations [6]. 
\s
\s
\n We shall now generate the metric (3.4) by a geometric ansatz [7]. Consider
5-Minkowski spacetime

$$ ds^2 = dt^2 - dx^2 - dy^2 - dz^2 - d w^2 \eqno (3.5)$$

\n and now impose the restriction

$$ x^2 + y^2 + z^2 + w^2 = K^2 (x^2 + y^2 + z^2) \eqno (3.6) $$

\n where $K^2 = (1 + 2k)^{-1} $. Elimination of the extra variable
leads to the metric (3.4). Note that the uniform $\phi$ metric
has only its space part curved which is described by the metric induced
on a cone, specified by the ansatz (3.6) in a 4-Euclidan space.
\s
\s
\n As demonstrated earlier curvature
of (3.4) is due to non-linear character of GR, which
persists even when $\phi$ is constant.  This 
is the remarkable and purely relativistic effect directly arising
from  non-linearity of the theory. Here ``constant $\phi$'' attains
physical meaning. This can be
seen geometrically as follows :  
\s
\s
\n Consider flat space 3-metric, $dl^2 = dr^2 + r^2 (d \theta^2 
+ sin^2 \theta d \varphi^2) $. The metric (3.4) corresponds to
$\phi = k $ which amounts to ``curving'' this metric by
introducing a const. $\not= 1 $ as a coefficient for $dr^2 $. 
It will now have curvature proportional to a sphere (isocurvature
spheres). The $\phi $ of (2.9) incorporates gravitational
acceleration in geodesic as well as ``curves'' 3-space.
The latter is, as argued earlier, the non-linear effect which
is not anulled out even when $\phi $ is constant. This is how constant
$\phi $ produces curvature and consequently attains physical meaning. 
\s
\s
\n For measuring this effect we should  do 
exactly what we do to measure curvature of a sphere. Let  two 
particles move on a $r = const.$ surface and then measure how  they 
come  closer.  Consider  a spaceship in an orbit  which  will  be 
freely  falling  and  hence free of gravity.  Let  two  particles 
propogate  across  the ship and the amount of  their  convergence 
will measure the curvature of the metric (3.2). That will measure the
constant $k $.
\s
\s
\n The curvature of spacetime goes as $r^{-2} $  and hence it
 will diverge as $r \longrightarrow 0 $.       
It makes curvature singular. The important point to  note  is 
how do we measure curvature? Only through the geodesic deviation, 
does  that  diverge? Remember that particles  need  to  propogate 
transversely because only $R^{23}_{~~23} $ alone is non-zero. At   
$r = 0 $,     there  cannot  be  any transverse motion.  It  
is  the  same 
situation as curvature of sphere diverges as $r \longrightarrow 0 $.
Hence it  is 
not a physically realisable singularity and is rather inocuous.
This is the asymptotic limit of the Schwarzschild global monopole
spacetime (3.2). Thus all physical measurements in (3.2) should be
referred to it.
\s
\s
\n The spacetime (3.4) could in some sense be thought
of as ``minimally'' curved because, (i) it has zero gravitational
mass $(\rho_c = 0)$, (ii) it is free of radial acceleration
and tidal acceleration occurs only for transverse motion and
(iii) its curvature is purely generated by constant relativistic
potential $\phi$.
\b

\item{\bf 4.} {\bf Discussion }
\s
\n We  have  demonstrated above that relativistic $\phi$, 
that specifies  the 
schwarzschild  field,  is determined absolutely by  the  Einstein 
vacuum  equations. That is, its non-zero constant value  has  
non-trivial physical  effect. This has happened  because  the  field 
equations also dictate the boundary condition leaving no  freedom 
to  relate  this  field  with any  other  at  the  boundary.  The 
spacetime has to be asymptotically flat if it were to be  vacuum. 
This is rather very extraordinary and unique feature. Usually one 
has freedom to choose boundary conditions to relate the situation 
under consideration with other field at the boundary.
\s
\s
\n In a real Universe an isolated body cannot be described  strictly 
by the Schwarzschild solution because the body does not exist all 
alone   in  the  Universe.  There  are  other  bodies. And gravity
can never be screened off.  A   good 
approximation  would be  that a body is sitting at the centre  of 
an empty spherical cavity surrounded by homogeneous and isotropic 
matter-energy  distribution,  representing rest of  the  Universe 
(ROU).  As  a matter principle the  Schwarzschild  solution  must 
admit  existence  of  non-empty  ROU  to  accord  with  the  real 
Universe. But it cannot as we have shown above because asymptotic 
flatness,  which cannot permit presence of any other  matter,  is 
inherent  in the solution. Giving up asymptotic flatness  implies 
giving up vacuum as well. In practice the Schwarzschild  solution 
is  very successful but we are raising a question   of  principle 
that it must be asymptotically non-flat  to be consistent 
with non-empty ROU. 
\s
\s
\n Can  this be achieved without disturbing the basic  character  of 
the  field so as to continue to enjoy the observational  support? 
The  answer is yes, as we have shown in  Section 3. That  is  the 
field  is  described  by a scalar function, as  before,  with 
restoration  of freedom to choose its zero. It is the  denial  of 
this  freedom  that led to asymptotic flatness.  Thus  the  field 
should  remain basically undisturbed for the key equation in  the 
problem is the Laplace equation that continued to hold good.  The 
additional  parameter $k$ in the generalized metric (3.2) will  now 
relate to the constant gravitational potential produced by ROU in 
the  interior  of the cavity. The picture now  becomes  identical 
with  the  Newtonian picture with the basic difference  that  the 
field is now sensitive to absolute value of the scalar function 
defining it. In NT, since field was not sensitive to absolute 
value of potential and hence constant potential  produced 
by  ROU in interior of cavity did not make  any  physical 
difference. In GR, on the other hand, non-zero constant potential 
due to non-empty ROU will now make a non-trivial contribution  to 
spacetime  in  the  interior  of the cavity. But it does not
alter the basic character of the field. This  is  the only and the
most harmless way to make the Schwarzschild field asymptotically
non-flat so as to be consistent with  the 
realistic setting.
\s
\s
\n It  is  remarkable  that the  above  proposed  generalization  is 
equivalent   to  ascribing  a  global  monopole  charge  to   the 
Schwarzschild particle. It is clearly demonstrated that existence 
of  non-vacuous  ROU is equivalent to adding  a  global  monopole 
charge.  The spacetime (3.4) generated by constant  $\phi$  is 
the  same as the global monopole metric with M = 0. This is  very 
interesting  and may be a manifestation of a deeper  relationship 
between  ROU  and  global monopole charge.  Note  that   constant 
potential  due to ROU is in fact the measure of global  monopole  charge. 
Global monopole is caused by spontaneous breaking of global  $O(3)$ 
symmetry  into  U(1) and is viewed as a topological  defect [4].  The 
other  thing to note is that this feature is purely  relativistic 
as  it  is shown in Section 2 that it arises  from  curvature  of 
space  part  of the metric which is  responsible  for  non-linear 
aspect  of  gravity (it being its own source). It  is  this  that 
lends  physical  meaning  to ``constant $\phi$''.  This  is  how 
constant  $\phi$  which  is physically  inert  in  NT  becomes 
physically active in GR.
\s
\s
\n What  we mean by potential here is a scalar function  that  fully 
determines  the field. It corresponds to the Newtonian  potential 
in  the  limit.  In the standard  curvature  coordinates,  we  do 
ultimately come to $R^0_0 = -\bigtriangledown^2 \phi = 0$,
hence it satisfies the  Laplace 
equation corresponding to $R^0_0 = 0$.
Any scalar function, not only being a solution of  the 
Laplace equation given by $R^0_0 = 0$, that fully determines the field 
and radial acceleration for free particle 
should  be  entitled to be termed potential.  In  either  theory, 
field of a body at rest can be described by only one scalar function.
The  point that disturbs people 
is  the  fact that the vacuum equations determine this scalar 
absolutely and so constant but non-zero value of it is physically 
distinguishable  from  the  zero value. This is  new  and  unique 
feature  of  GR  and  is  directly  related  to  its   non-linear 
character. 
\s
\s
\n Leaving the question aside for the moment whether $\phi$
in (2.9) represents potential in GR or not, it is clear that it
completely describes the field of a Schwarzschild particle. 
Irrespective of whether it is NT or GR, field of a mass point can
require no more than one quantity to describe it. All what we say
is that the Einstein vacuum equations determine this quantity
absolutely while NT offers freedom to choose its zero arbitrarily.
In order to incorporate the feature that gravitational field
energy has gravitational charge, the Newtonian Laplace equation
needs to be modified to take the non-linear form (1.1). This would
however be referred to flat spacetime background. In GR, on the other
hand background spacetime is curved that permits us to retain
the Laplace equation as it is with non-linear aspect being taken 
care of by ``curving'' space. The price we pay for it is that 
$\phi$ now gets determined absolutely. This happens because
now $\phi$ has also to ``curve'' space appropriately
to exactly counteract field energy density on the right of (1.1).
This it could do only by fixing one of the two free parameters
$k$ and $M$ in (2.9). $M$ cannot be specified as it represents
mass of the body that should remain free. Thus $k$, which is
physically inert in classical physics, can only be fixed.
This is how $\phi$, the solution of the Laplace equation is
determined absolutely in GR. Its absolute zero value is  given by
asymptotic flat spacetime. This is the trade off for incorporating
the distinguishing non-linear aspect of the theory.
\s
\s
\n Finally  we  wish  to say that the Einstein vacuum equations determine
the relativistic $\phi$ absolutely and its zero being defined
by the asymptotic flat spacetime. Retaining the free parameter $k$
in (2.9) is  the  only  way  to 
generalize  the Schwarzschild field to render  it  asymptotically 
non-flat  to  be  in  principle  consistent  with  the  realistic 
setting  as  it actually obtains in the Universe. At  the  same 
time  the basic physical character of the field that  has  strong 
observational support is not significantly disturbed. 
\s
\s
\n{\bf Acknowledgement :}
\s
\n Over a period of past one year I have benefitted from discussions
and criticism from several friends, and they included amongst others
Sailo Mukherjee, Jayant Narlikar, Reza Tavakol, Jose Senovilla 
and Malcolm MacCallum. This however should not be taken to mean
that they all share  this view.
Their criticism was nonetheless very useful for me to gain insight and
sharpen some of the arguments and I hope that they would
perhaps be happier with the final version. I thank 
them all warmly.
\vfill\eject
\c{\bf\mid References :}
\s
\item{1.} N. Dadhich, GR-14 Abstracts, A.98 (1995).
\s
\item{2.} H. Stephani, General Relativity (Cambridge University
Press, 1990), p.99.
\s
\item{3.} R.A. D'Inverno, Introducing Einstein's Relativity (Oxford
University Press, 1992) p.187.
\s
\item{4.} M. Barriola and A. Vilenkin, Phys. Rev. Lett.,
{\bf 63}, 341 (1989).
\s
\item{5.} N. Dadhich, K. Narayan and U. Yajnik, Schwarzschild black hole
with global monopole charge, submitted.
\s
\item{6.} N. Dadhich, Ph.D. thesis (Poona University, 1970) unpublished.
\s
\s
\item{7.} N. Dadhich and K. Narayan, An ansatz for spacetimes of zero
gravitational mass : global monopoles and textures, submitted.

\bye